# Compressive Imaging with Stochastic Spatial Light Modulator

J.C. S CHAAKE,[1,2]* R.C. P OOSER,[2] S. J ESSE[3]

[1]Department of Physics and Astronomy, The University of Tennessee, Knoxville, Tennessee 37996-1200, USA
[2]Quantum Information Science Group, Oak Ridge National Laboratory, Oak Ridge, Tennessee 37831, USA
[3]The Center for Nanophase Materials Sciences, Oak Ridge National Laboratory, Oak Ridge, Tennessee 37831, USA
*Corresponding author: jschaake@vols.utk.edu

**We present a stochastic analog spatial light modulator designed for compressive imaging applications. We rely on the unpredictable nature of multi-particle collisions to provide randomization for the particle location. We demonstrate this concept in an optical imaging system using a single-pixel camera. This design can be applied to imaging or spectroscopic systems in which no analog to optical spatial light modulators currently exist or in non-optical lensless imaging systems.**

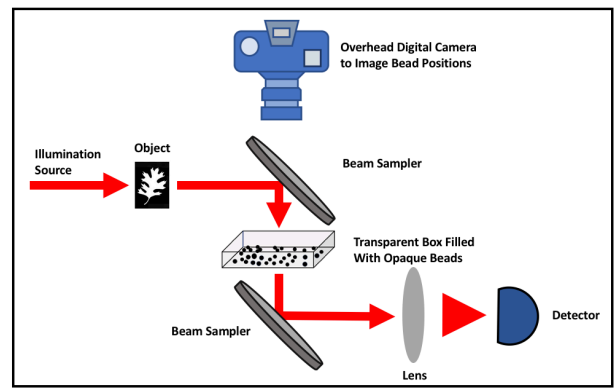

**Fig. 1.** Schematic of our compressive imaging setup. The object to be imaged is illuminated. The image is sampled using an analog spatial light modulator. A pair of beam samplers are used to create a periscope that passes light vertically through the box of particles. The sampled light is collected by a lens and focused onto a single detector element. A digital camera is used to simultaneously image the positions of the random array of particles from above. The sampled light power gathered, along with each sampling array, are both stored for processing. The modulator is reconfigured and the process is repeated.

Sparse sampling or compressive sensing (CS) is a technique for reconstructing a signal, that is sparse in some basis, using fewer samples than required by classical information theory [1]. In the case of imaging with optical fields, samples of the signal are taken using a basis that is minimally coherent with the signal. Sampling is performed by imprinting the sampling matrix upon the signal using a spatial light modulator (SLM). A SLM such as a digital micro-mirror device (DMD) [2][3], or a liquid crystal device (LCD) [4], [5] [6], or a photonic crystal slab [7] can all be used to sample the signal. This has led to so called "lensless cameras" [4], [21] employing CS.

The sampled signal is collected using a single detector element. This analog signal is converted to a digital detector signal that is stored along with each corresponding mask. The sets of detected signals along with the sampling masks are passed to reconstruction algorithms. These algorithms are typically $l_1$-minimization routines for finding solutions to systems of underdetermined linear equations. These techniques have become a useful tool for applications such as image denoising [8] [9], seismic data handling [10], image reconstruction in the millimeter wavelengths [11], infra-red wavelengths [12], X-ray wavelengths [13][14], real time object tracking [15], laser ranging [16], sub-shot-noise limited imaging [17], single-photon imaging [18]–[20], and diagnosing optical beam characteristics down to the single-photon level [3].

We have devised a stochastic spatial light modulator (Fig. 1) which will allow for the characterization of systems that currently have no optical counterpart. This design employs a small, optically transparent container filled with particles. These particles are opaque to the radiation being used for imaging. Whereas traditional imaging systems require arrays of detectors to generate an image, compressive imaging (CI) systems can employ a single detector element. This can greatly reduce the electronics noise in the system. Standard CI systems employ a digital SLM which must be compatible with the radiation being studied. These are readily available for optical wavelengths ranging from X-ray [13, 14], to IR [3, 12]. CI is useful for studying systems in which arrays of detectors are cost prohibitive or do not exist. Our design furthers this by

Notice: This manuscript has been authored by UT-Battelle, LLC, under contract DE-AC05-00OR22725 with the US Department of Energy (DOE). The US government retains and the publisher, by accepting the article for publication, acknowledges that the US government retains a nonexclusive, paid-up, irrevocable, worldwide license to publish or reproduce the published form of this manuscript, or allow others to do so, for US government purposes. DOE will provide public access to these results of federally sponsored research in accordance with the DOE Public Access Plan (http://energy.gov/downloads/doe-public-access-plan).

removing the controlled SLM and replacing it with a randomized array of particles. Exact control over the SLM is not necessary if one can simply determine the position of the absorbing particles. We accomplish this by capturing an image of the particles at the time of measurement using a digital camera.

The particles we began with, were 3 mm external diameter, torus shaped, black, glass beads. The image to be sampled was sent vertically through the container using a periscope created using a pair of glass beam samplers (Fig. 1). This geometry allowed for the particles to be randomized using a small electric motor.

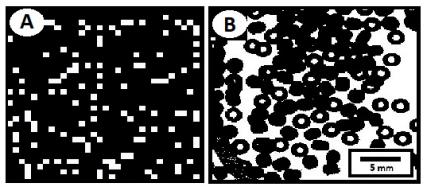

**Fig. 2.** Figure A is an example of a standard random matrix that is sent to, for instance, a DMD. Figure B is a random matrix derived from an image of the random SLM with a threshold applied.

After disturbing the container, a small amount of time was allowed for the particles to come to rest. This vibration randomizes the positions of the particles, resulting in a random, or stochastic "sampling matrix". A random matrix is minimally coherent with most signals, so this is an ideal method for generating sampling matrices. To generate the sampling matrix element, the container was imaged from above. The image was trimmed to the container dimensions and a threshold was applied to each pixel of the grayscale image using a simple Matlab routine. This results in a black and white sampling matrix similar to a standard SLM (Fig. 2). Whereas a standard sampling matrix is generated using computer code (Fig. 2(A)), we rely upon the complex interaction of numerous bodies colliding within a closed container to randomize the position of the particles (Fig. 2(B)). It may be noted, in the images in Figs. 2 and 3, that the particles tend to cluster in the center of the image. This is due slightly to static charge buildup, that clings the particles together and also to the chamber walls. The predominant driver of clustering is the vibration of the chamber itself. An electric toothbrush motor was used to drive the vibrations. These devices generate a cyclic motion which is translated to the particle chamber. A larger chamber along with a more powerful motor would reduce these effects.

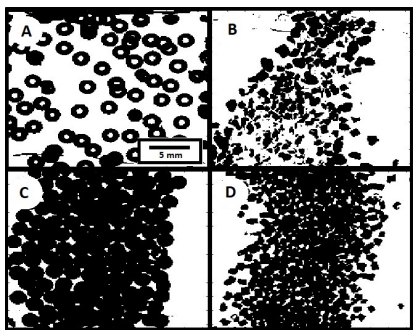

**Fig. 3.** Above are examples of sampling matrices using a variety of particles. (A) is an image of glass beads. (B) is an image of modeling sand, (C) is an image of 3 mm diameter, flat, opaque disks. (D) is an image of dyed black sugar crystals.

An image of the random SLM was captured simultaneously with a measurement of the intensity of light at the power meter. The image and the power reading were saved in individual indexed arrays for later processing. This pair of matrices was sent to a reconstruction algorithm. In this case, a Total Variation method, TVAL [22], was used for image reconstruction. This algorithm was chosen because of rapid convergence for imaging applications. The user controls the number of samples collected, the time that the motor randomizes the particle array, and the threshold level for image processing. User also is free to choose the particles used in the array. The particles only need to be opaque to the radiation used for image formation.

Initially, reconstructions were performed with a simple Gaussian laser beam to demonstrate system operation. A 5mm FWHM 790 nm CW laser beam was passed through the cell and collected by the power meter. Imaging was performed with a variety of particles, and sizes, to demonstrate the robustness of this design. Examples of the sampling matrices generated from four different particles are shown (Fig. 3): opaque glass beads 3(A), modeling sand 3(B), opaque, flat disks 3(C), and dyed, black sugar crystals 3(D). All particles functioned appropriately but the smaller, and lighter particles were prone to static buildup during runs. This clumping reduces the randomness of the sampling elements. As randomness is a key aspect to this sampling method, any reduction in randomness causes poorer image reconstruction.

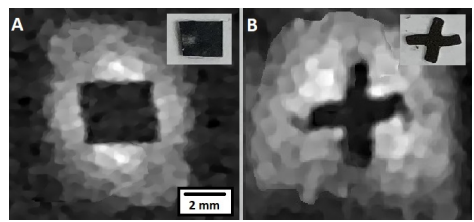

**Fig. 4.** Various images reconstructed to demonstrate the spatial mode definition provided. Image A is a 3 mm x 5 mm rectangle. Image B is a cross with dimensions of 6 mm x 6 mm with arms 1 mm wide. Actual objects imaged are presented in each inset. Image rotation is due to slight misalignment of object and camera.

Images were generated using a variety of masks (Fig. 4). A small 3 mm by 5 mm rectangle was imaged (Fig. 4(A)), along with a "cross" (Fig. 4(B)) with 9 mm by 9 mm with 1 mm wide arms. Each mask was illuminated by an expanded Gaussian beam from our 790 nm laser. Black glass beads were used as the absorber in both cases. The image values generated by the camera are grayscale and range between 0 (black) and 100 (white). A user defined threshold point converts the image to a black and white (digital) image. For our purposes, selecting an appropriate threshold level reduces light scattered from the particle surfaces and transmitted light in cases of sand and sugar. The effect of thresholding level on image quality is seen in the quality of the images generated (Fig. 5). A comparison of two different absorbers shows the importance of threshold level for each particle. Figures 5(A-C) were generated using dyed decorative sugar granules. Figures 5(D-F) were generated using black glass beads. The threshold levels applied in each image are: 90% 5(A) and 5(D), 60% in 5(B) and 5(E), and 30% in 5(C) and 5(F). Examples of each sampling matrix are shown in the upper right inset of each image.

The highest threshold level results in the largest assumed area of absorption. As the threshold level was reduced to 60% (see: 5(B) and 5(E)), the image of the absorbers more appropriately matches the true absorber array and the imaging is optimal. As the threshold level was further reduced (see: 5(C) and 5(F)), we assumed that the particle edges transmit the image well instead of absorbing. This means that the sampling matrix no longer matches what is happening in the cell and the image again becomes blurred.

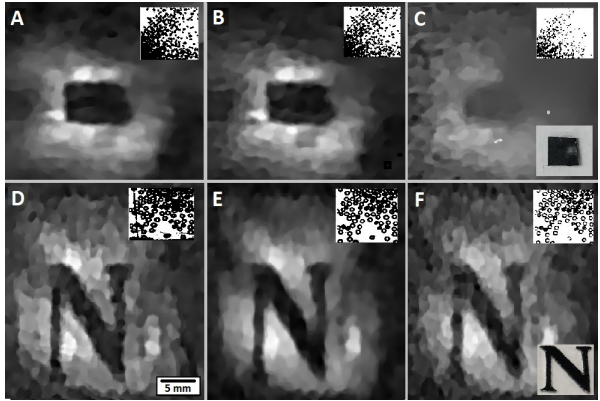

**Fig. 5.** Examples of images generated using varying threshold levels and two different particles. Figures (A), (B) and (C) were generated using sugar granules. Figures (D), (E), and (F) were generated using glass beads. The threshold levels used are: 90% (A) and (D), 60% (B) and (E), and 30% (C) and (F). An example of the sampling matrix with each threshold applied is shown in the top, right corner of each reconstruction. The actual masks used for imaging is shown in the bottom inset in (C) and (F).

With a traditional compressive imaging setup, the resolution (R) is given by:

$$R = \left(\frac{Total\ Image\ Width\ (in\ pixels)}{Super\ Pixel\ Size}\right)^2 \quad (1)$$

Total image width is the width of the region on the SLM used for sampling. If the individual elements of the SLM are much smaller than the feature being sampled, pixels are grouped together creating a "super-pixel". By increasing the super-pixel size, one can greatly reduce both sampling and computational time at a cost of decreased resolution. As our system does not have a true "pixel" size, it would be useful to calculate an effective resolution. As our digital camera captures a 250 x 250-pixel image of the 25 mm$^2$ transparent container, we can immediately begin with a maximum resolution of 250$^2$, or 62,500 pixels/mm provided by the digital camera. Since we are using particles to sample, there should be some "effective super-pixel" size. For example, if we began with the small 2.97 mm$^2$ beads, each bead corresponds to a 17.2-pixel element. Using this as our "effective super-pixel" size, we derived a resolution of 216 pixels/mm. It should be clear from the sample images of the arrays used for testing that the random arrays of particles do not function as individual elements like a standard DMD making it more difficult to determine the resolution limits of this imaging method.

Increasing the number of samples used for reconstruction increases the resolution of the image. This is demonstrated in Figure 6. Initially, an image (Fig. 6(A)) was generated using 1000 samples. When the sample number was dropped to 500 (Fig. 6(B)), the circular features due to the particle shape are no longer averaged out and began to show up more noticeably. The rectangle was still clearly visible. Dropping the sample number to 250 (Fig. 6(C)) seriously degrades the image. The overall shape was still visible but was not clearly discernable. Any edge features have been lost. Reducing the sample to 100 (Fig. 6(D)) removes the feature almost completely.

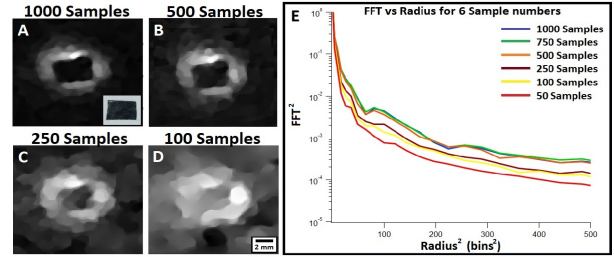

**Fig. 6.** Images are reconstructed using decreasing number of samples. (A) was reconstructed using 1000 samples. The actual image is shown in the inset. (B) the sample number is reduced to 500 with slight degradation in image quality. (C) the sample number is reduced to 250 with serious degradation in image quality. (D) the sample number is reduced to 100 samples; the image is no longer discernable. (E) is a plot of each sample number. The 2-d FFT of each image, radially integrated, is plotted vs the radius. The decreased FFT value shows a lower cutoff frequency for the image and thus decreased resolution.

More systematic characterizations of the resolving limits of this imaging method were carried out. Taking the 2-d Fast Fourier Transform (FFT) and radially integrating gives a function which demonstrates the frequency cutoff of the image (Fig. 6(E)). This function drops steeply, then has a large noise floor, before dropping to zero at high frequencies. To derive a resolution limit, we look at the steep portion of the curve. While not linear, steeper slopes for this curve indicate lower cutoff frequencies. Six different sample numbers were plotted. For 1000, 750, and 500 samples, the function is very similar. 250 samples show a slight decrease in value, which corresponds to a decrease in cutoff frequency. Further reductions to 100, and 50 samples show greater reductions cut off frequency as expected.

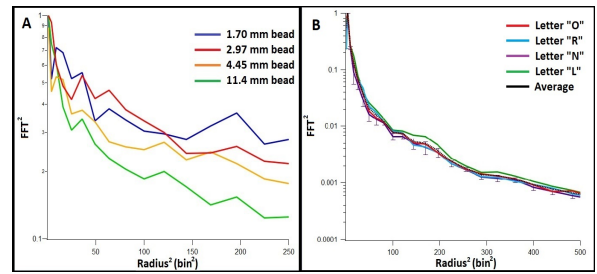

**Fig. 7.** Plots demonstrating the resolving limit of our system. Fig. (A) contains plots of the 2-d FFT vs radius for 4 different particle sizes. As the particle size increases, the slope of the steep, initial portion of each function (frequency cutoff) decreases, showing reduced resolution. Fig (B) is a plot of the FFT vs radius generated from four different images created using the same particle and sample number. The error bars for the average value of the four functions (black line) contain all but a small portion of the functions demonstrating reliability of the design.

The area of the absorbing particle was should have an effect on the resolution. The 2-d FFT was again calculated for an image generated using four different sized particles. This FFT was radially integrated and plotted as a function of radius. The resulting curves shown in Figure 7(A) demonstrate that the resolution drops as the bead size increases. In Figure 7(B), we plot the 2-d FFT for four different images generated using the same particle and same number of samples. These plots should be identical as the resolution is driven by particle size and sample number. Error bars on the curve produced by averaging the runs show that the curves overlap for all but a small portion of the function. This demonstrates that calculating the resolution limits of the system using this method is reliable.

Other more complex images were created using the 2.97 mm$^2$ particles (Fig. 8 "ORNL" logo and name). The actual masks used for imaging are shown below. The very fine structure of the oak leaf is not resolved, but the overall shape is defined. The larger letters are resolvable but not the smallest letters at the bottom of the image. The mask used to create the image was larger than the cell, so the image was taken in two steps and added together. For larger images generated with this setup, the epoxy used to attach the container to the motor causes severe scattering and blurs the image edges. Also, edge of the glass microscope slide used in the periscope, to reflect light to the detector, is visible.

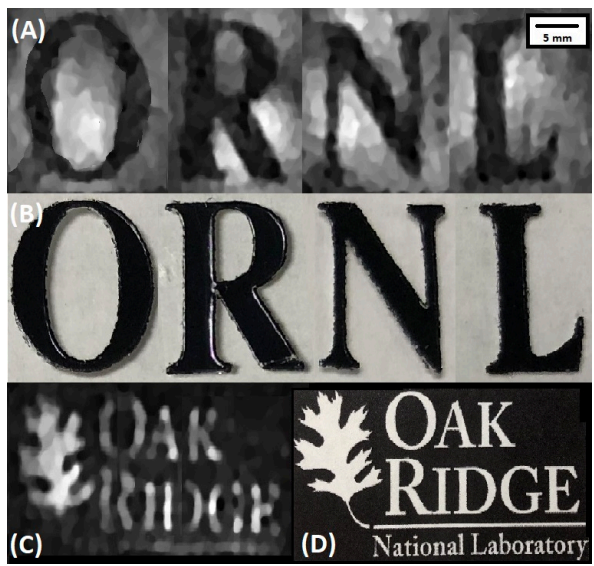

**Fig. 8.** A demonstration of the imaging capabilities of our system. (A): the "ORNL" letters were each imaged individually and combined into a composite image. (B): The masks used for imaging. (C): the "Oak Ridge National Laboratory" logo was imaged. The mask used in seen in (D). Reconstructions were performed using the 2.97 mm$^2$ beads. Mask was illuminated using an expanded 790 mm laser beam.

We have demonstrated an analog, stochastic compressive imaging system. A small, transparent enclosure is filled with a number of opaque particles. The enclosure is mounted on a vibrating platform, that randomizes the particle positions. For each sample, the particles are randomized, and an image of the array is taken, along with a power reading of the transmitted light. A threshold is applied to the image of the array to create a "sampling matrix". The set of sampling matrices and power data are sent to a TVAL optimization algorithm, creating an image. The system was tested using a variety of particles, from glass beads to dyed sugar crystals. Resolving capabilities were tested by generating an image, then taking the 2-d FFT, radially integrating, and plotting as FFT value as a function of radius. The functions generated show a steep drop in value initially, flatten out across the noise floor, then drop to zero. The slope of the steep initial portion of the curve (plotted) gives the cutoff frequency. A steeper slope means a lower cutoff frequency and reduced resolution. The resolution of our system depends on the number of samples used for reconstruction. Using larger sample numbers gives greater resolution. This measurement was repeated using 4 different particle sizes to demonstrate that resolution also scales with particle size. We also plotted multiple functions generated using the same particle and sample size. The functions generated were all consistent demonstrating robustness of our design.

This design is flexible and gives the user the capability to image systems which have no current spatial light modulator available.

This work was performed at Oak Ridge National Laboratory, operated by UT-Battelle for the U.S. Department of Energy under contract no. DE-AC05-00OR22725. Funding was provided by the Intelligence Community post-doctoral fellowship program.